\documentclass[
aps,
prc,
superscriptaddress,
floatfix,
nofootinbib,
twocolumn,
]{revtex4-2}

\usepackage{amsmath,amssymb,amsfonts,physics,graphicx,float}
\usepackage{stmaryrd}
\graphicspath{{figs/}}
\usepackage{xcolor}
\usepackage{ulem}
\usepackage[setpagesize=false]{hyperref}
\usepackage{chngcntr}

\allowdisplaybreaks[4]

\usepackage{color}
\definecolor{ar}{rgb}{1.0, 0.01, 0.24}
\definecolor{al}{rgb}{0.82, 0.1, 0.26}
\definecolor{ev}{rgb}{0.56, 0.0, 1.0}

\definecolor{brn}{rgb}{0.45, 0.24, 0.09}
\definecolor{BRN}{rgb}{0.45, 0.24, 0.09}%

\def\be{\begin{eqnarray}}
\def\ee{\end{eqnarray}}

\begin{document}

\author{Yao Ma}
\email{mayao@nju.edu.cn}
\affiliation{School of Frontier Sciences, Nanjing University, Suzhou 215163, China}

\author{Bikai Gao}
\email{bikai@rcnp.osaka-u.ac.jp}
\affiliation{Research Center for Nuclear Physics (RCNP), Osaka University, Osaka 567-0047, Japan}

\author{Masayasu Harada}
\email{harada@hken.phys.nagoya-u.ac.jp}
\affiliation{Kobayashi-Maskawa Institute for the Origin of Particles and the Universe, Nagoya University, Nagoya 464-8602, Japan}
\affiliation{Department of Physics, Nagoya University, Nagoya 464-8602, Japan}
\affiliation{Advanced Science Research Center, Japan Atomic Energy Agency, Tokai 319-1195, Japan}

\author{Yong-Liang Ma}
\email{ylma@nju.edu.cn}
\affiliation{School of Physics, Nanjing University, Nanjing 210093, China}
\affiliation{School of Frontier Sciences, Nanjing University, Suzhou 215163, China}

\title{Chiral invariant mass and scale symmetry in dense matter from a dilaton compensated parity doublet model}

\date{\today}

\begin{abstract}
In the parity doublet model the nucleon mass includes a chiral invariant part $m_0$, which is conventionally assumed to be medium independent.
We promote this assumption to a dynamical question by making the model scale invariant using a dilaton compensator $\chi$, so that the effective chiral invariant mass $m_{\rm inv}^\ast = m_0\tilde\chi$, with $\tilde\chi = \chi/f_\chi$, responds to the partial restoration of scale symmetry in the medium.
We fit the model parameters by using vacuum spectra, nuclear matter properties, and NICER mass--radius posteriors---with the hadronic equation of state connected to 2+1-flavor Nambu--Jona-Lasinio quark matter by an interpolation.
The best fit gives $m_0 = 657$~MeV, a maximum neutron-star mass of $2.30\,M_\odot$ and $R_{1.4} = 12.22$~km, with the mass--radius sequence passing through all three NICER credible regions. The resulting tidal deformabilities, $\Lambda_{1.4} = 387$ and $\tilde\Lambda = 463$, are consistent with GW170817 measurements.
The in-medium analysis separates the two restorations: at the hadron--quark matching density the chiral condensate falls to $56\%$ of its vacuum value, while the dilaton decreases only to $94\%$, so the nucleon arrives at the matching density with $90\%$ of its mass carried by the invariant part.
The order parameters of chiral and scale symmetries organize the in-medium probes, making the $\omega$ an almost pure probe of scale symmetry and the $\rho$ of chiral restoration, which gives us a clue on the origin of hadron mass.
\end{abstract}

\maketitle

\section{Introduction}
\label{sec:intro}

Understanding the origin of the nucleon mass and how it rearranges in dense matter are central questions of hadron physics and neutron star (NS) phenomenology. In the conventional linear sigma model the nucleon mass is entirely generated by the spontaneous breaking of chiral symmetry, and would vanish when chiral symmetry is restored. The parity doublet model (PDM)~\cite{Detar:1988kn,Jido:2001nt} offers a structurally different possibility: the positive-parity nucleon $N(939)$ and its negative-parity partner, often taken as $N^\ast(1535)$, are placed in the same chiral multiplet with mirror assignment. This assignment allows a chiral invariant mass $m_0$---a piece of the baryon mass that survives after chiral symmetry restoration---so that {two nucleons} become degenerate at finite mass $m_0$ rather than massless. Lattice simulations at finite temperature~\cite{Aarts:2015mma,Aarts:2017rrl,Aarts:2018glk} and QCD sum rule analyses~\cite{Kim:2020zae,Kim:2021xyp,Lee:2023ofg} support the parity doublet structure qualitatively. Since the pioneering mean-field construction~\cite{Hatsuda:1988mv}, the PDM has become a useful framework for investigating the origin of nucleon mass. In particular, the analysis of nuclear matter (NM) and NS typically constrains $m_0$ to a broad window of roughly $500$--$900$~MeV~\cite{Zschiesche:2006zj,Motohiro:2015taa,Minamikawa:2020jfj,Marczenko:2018jui,Gao:2022klm}. In addition, the model was also developed in several directions. Recent extensions include an explicit isovector-scalar channel~\cite{Kong:2023nue}, three-flavor parity doubling~\cite{Fraga:2023wtd}, thermodynamically consistent extensions of the standard mean-field treatment of the doublet~\cite{Eser:2023oii,Eser:2024xil,Recchi:2025pyy,Recchi:2026sgo}, a microscopic quark--diquark realization in which both the mirror assignment and the invariant mass itself emerge from diquark dynamics~\cite{Gao:2026jsg}, hybrid-star constructions with large quark cores~\cite{Yuan:2025dft}, accelerated NS cooling through direct-Urca processes involving the negative-parity partner~\cite{Brodie:2025nww,Negreiros:2026ode}, confrontations with multi-messenger data ranging from the light compact object in HESS J1731$-$347 to the newest NICER radii~\cite{Marczenko:2021uaj,Gao:2024chh,Gao:2025vdc}, constraints on the invariant mass obtained with a model-independent treatment of the intermediate-density region~\cite{Gao:2025nkg}, and an extraction of the chiral invariant mass from the nucleon gravitational $D$-form factor~\cite{Kawaguchi:2025cuf}.

In the above studies the chiral invariant mass is a vacuum quantity that does not depend on medium.
That is a strong assumption, because the dynamics, from which the invariant mass arises, is not medium blind.
For example, Ref.~\cite{Gao:2026jsg} attributes $m_0$ to the colored diquark inside the baryon---both its mass and its binding to the remaining quark are supplied by non-Abelian gluon dynamics rather than the quark condensate, which is precisely why the generated mass survives chiral restoration.
We note that the same gluon dynamics is what breaks scale symmetry: the gluonic contribution to the baryon mass, now being quantified on the lattice through the trace anomaly~\cite{Metz:2020vxd,Hu:2024mas}, does not need to be unchanged.
If $m_0$ encodes the gluonic contribution, it should be responsible for the partial restoration of scale symmetry in a dense medium, in the spirit of Brown--Rho scaling~\cite{Brown:1991kk} and of the pseudo-conformal picture of dense matter~\cite{Ma:2018jze,Rho:2021zwm}. The natural framework in which this response can be formulated is a scale-invariant Lagrangian built with a dilaton field $\chi$, the (pseudo-)Nambu--Goldstone boson of the spontaneously (and anomalously) broken scale symmetry. In this study, we construct the scale-invariant PDM (sPDM) by dressing every dimensional operator with the appropriate power of $\chi/f_\chi$, where $f_\chi$ is the dilaton condensate. In particular the chiral invariant mass term becomes
\begin{equation}
m_0\, \bar\psi \psi \;\longrightarrow\; m_0\, \frac{\chi}{f_\chi}\, \bar\psi \psi\,,
\end{equation}
so that the ``chiral invariant'' part of the nucleon mass is no longer constant but changes with the medium through the dilaton mean field. The Lagrangian parameter $m_0$ itself remains a constant, and what acquires a density dependence is the effective invariant mass $m_{\rm inv}^\ast \equiv m_0\tilde\chi$, with $\tilde\chi \equiv \chi/f_\chi$. The scale-dressed
chiral invariant mass $m_0\tilde\chi$ has appeared before---in the dilaton-limit fixed-point analyses of Refs.~\cite{Sasaki:2011ff,Paeng:2011hy} and, as a scalar-generated invariant mass $m_0 = g_\chi\chi$, in the hybrid quark--meson--nucleon model of Ref.~\cite{Marczenko:2017huu}. What is new here is the integrated construction: the Crewther--Tunstall potential with the slope $\beta'$ of the QCD $\beta$ function at the infrared fixed point---the anomalous dimension of the gluonic operator $G^2$, the scale-dressed chiral invariant mass in a hidden local symmetry (HLS) parity doublet model, and a controlled matching to a Nambu--Jona-Lasinio (NJL) quark equation of state (EoS).

In our construction, the in-medium nucleon mass matrix contains two pieces of distinct symmetry origin. One piece is proportional to the scalar mean field $\sigma$ which vanishes when chiral symmetry is restored; the other is $m_{\rm inv}^\ast = m_0\tilde\chi$ which tracks the scale-symmetry order parameter, and the physical mass combines both through Eq.~\eqref{eq:rotated}. Correspondingly, the model introduces two distinct mean-field order parameters---$\Phi_\pi = \langle\sigma\rangle^\ast/f_\pi$ for chiral symmetry and $\Phi_\chi = \langle\chi\rangle^\ast/f_\chi$ for scale symmetry---whose density dependence characterizes the interplay between the chiral and scale symmetry breaking. 

In our model, the parameters are fixed through a global $\chi^2$ fitting. The fitted targets are vacuum hadron properties (nucleon and $N^\ast(1535)$ masses, $m_\rho$, $m_\omega$ and broad priors on the scalar masses), NM properties at the saturation density (the binding energy per nucleon $B/A$, the incompressibility $K_0$, the symmetry energy $E_{\rm sym}$ with its slope parameter $L$, and skewness parameter $J_0$), and NS observables (Neutron Star Interior Composition Explorer (NICER) mass--radius posteriors for PSR~J0030+0451, PSR~J0740+6620, and PSR~J0437$-$4715~\cite{Riley:2019yda,Miller:2019cac,Riley:2021pdl,Miller:2021qha,Salmi:2024aum,Choudhury:2024xbk}). The GW170817 tidal constraints---the binary deformability $\tilde\Lambda = 300^{+420}_{-230}$ at $90\%$ credibility (highest-posterior-density interval, low-spin prior)~\cite{LIGOScientific:2018hze}, and $\Lambda_{1.4} = 190^{+390}_{-120}$ at $90\%$ from the common-equation-of-state analysis of the same event~\cite{LIGOScientific:2018cki}---are checked a posteriori.
After fixing the sPDM parameters, the hadronic matter is connected to a 2+1-flavor NJL quark matter EoS at several times saturation density, by a direct interpolation of the energy density between a lower matching density $n_L$ and an upper density $n_H$ (referred to below as the ``polytrope'' connection).

The paper is organized as follows. Section~\ref{sec:model} presents the scale-invariant Lagrangian, the decomposition of the in-medium masses, and the two mean-field order parameters.
Section~\ref{sec:fit} details the fitting procedure, the objective, and the hadron--quark connection, then presents the {fit}, sets it against the frozen-dilaton $\sigma^6$-PDM, and examines the in-medium masses and couplings that dilaton controls.
Section~\ref{sec:summary} is a summary and outlook.

\section{Scale-invariant parity doublet model}
\label{sec:model}

\subsection{Lagrangian and mean-field equations}
\label{sec:lagrangian}

We consider a PDM in which scale symmetry is incorporated through a dilaton field $\chi$, and the $\rho$ and $\omega$ mesons are introduced as gauge bosons of the HLS~\cite{Bando:1987br,Harada:2003jx}, building on the framework of Ref.~\cite{Motohiro:2015taa}. The Lagrangian reads
\begin{equation}
{\mathcal L} = {\mathcal L}_S + {\mathcal L}_V + {\mathcal L}_{SV} + {\mathcal L}_N\,,
\label{eq:Ltot}
\end{equation}
 whose four terms describe the scalar and dilaton fields, the vector mesons, the couplings between the scalar and vector sectors, and the nucleons, respectively.

\paragraph*{Chiral and dilaton fields.}
{The meson field is introduced as}
\begin{equation}
M = \sigma + 2 i \sum_{a=1}^{3} \pi_a T_a\,,
\label{eq:Mfield}
\end{equation}
where $T_a$ is the Pauli matrix normalized to $1/2$, $\sigma$ and $\pi_a$ are the scalar and pseudoscalar fields which transform as $M \to g_L M g_R^\dagger$ under $\mathrm{SU}(2)_L \times \mathrm{SU}(2)_R$. Scale symmetry is carried by the dilaton {field} $\chi$. We introduce the dimensionless compensator $\tilde\chi \equiv \chi/f_\chi$, with $f_\chi \equiv \langle\chi\rangle_0$ denoting the vacuum expectation value of the dilaton field, such that $\langle\tilde\chi\rangle_0 = 1$ in vacuum. The scalar--dilaton Lagrangian is
\begin{align}
{\mathcal L}_{S}=&\frac{1}{4} \tr\left[ \partial^\mu M^\dag \partial_\mu M \right] + \frac{1}{2} \partial_\mu\chi\partial^\mu\chi
- V\!\left( M, \chi \right),
\label{eq:LS}
\end{align}
with the potential
\begin{align}
V \left( M,\chi \right) = & -
\frac{1}{4} \bar{\mu}^2 \tilde\chi^2 \tr\left[ M^\dag M \right]
+ \frac{1}{8} \lambda \tr\left[ M^\dag M M^\dag M \right]
\notag\\
& + \frac{h_4}{4} \tilde\chi^4 + \frac{h_5}{4 + \beta'} \tilde\chi^{4+\beta'}.
\label{eq:potential}
\end{align}
Here $\bar\mu^2, \lambda, h_4$ and $h_5$ are parameters in the potential. The powers of $\tilde\chi$ are fixed by scale invariance, so that the $h_5$ term provides the only explicit scale-symmetry breaking in the mesonic potential. The exponent of the $h_5$ term is governed by $\beta'>0$, the slope of the $\beta$ function at the infrared fixed point of the Crewther--Tunstall scenario~\cite{Crewther:2013vea}.

Denoting $\langle\sigma\rangle_0 = f_\pi$ and $\langle\chi\rangle_0 = f_\chi$, the vacuum stationary conditions
\begin{align}
& \frac{\partial V}{\partial \sigma}\bigg|_{\rm vac} = \frac{\partial V}{\partial \chi}\bigg|_{\rm vac} = 0
\notag\\
& \quad\Longrightarrow\quad
\bar\mu^2 = \lambda f_\pi^2\,, \qquad h_4 + h_5 = \bar\mu^2 f_\pi^2\,,
\label{eq:vacstat}
\end{align}
determine $\bar\mu^2 = (h_4+h_5)/f_\pi^2$ and $\lambda = (h_4+h_5)/f_\pi^4$, so that $h_4$ and $h_5$ both remain free parameters.

The potential contains no explicit chiral-symmetry-breaking term, so that the present analysis is carried out in the chiral limit, with a massless pion and the physical value of $f_\pi$ taken as input. Restoring the scale-dressed explicit chiral symmetry breaking and a finite $m_\pi$, is left for future work. In the frozen-dilaton limit used for comparison in Sec.~\ref{sec:sigma6} the dilaton is set to $\tilde\chi \equiv 1$.

\paragraph*{Hidden local symmetry and the vector sector.}
To introduce the vector meson fields, we make the decomposition
\begin{equation}
M = \xi_L^\dag\, \sigma\, \xi_R\,,
\label{eq:Mdecomp}
\end{equation}
where $\xi_{L,R}$ are the coset fields carrying pions and $\sigma$ is the scalar 
meson field. Using this decomposition, one can introduce a redundant local flavor symmetry $h(x)$ such that the relevant quantities have the following transformations
\begin{equation}
\xi_{L,R} \to h(x)\xi_{L,R} g_{L,R}^\dagger,
\end{equation}
with $\sigma$ being the singlet of $h(x)$. Corresponding to the local $h(x)$ symmetry, we introduce the HLS gauge field
\begin{equation}
V_\mu = g_\rho \sum_{a=1}^{3} \rho_\mu^a T_a + \tfrac{1}{2}\, g_\omega\, \omega_\mu\,,
\label{eq:Vmu}
\end{equation}
with $\rho_\mu^a$ and $\omega_\mu$ the isovector and isoscalar vector-meson fields ($\omega_\mu$ multiplying the unit matrix in isospin space) and $g_\rho$, $g_\omega$ the corresponding HLS gauge couplings. The HLS gauge field strength is given by 
\begin{align}
V_{\mu\nu} & = \partial_\mu V_\nu - \partial_\nu V_\mu - i \left[ V_\mu \,,\, V_\nu \right].
\end{align}
The covariant derivatives of the coset fields
are
\begin{align}
D_\mu \xi_R & = \partial_\mu \xi_R - i V_\mu \xi_R + i \xi_R {\mathcal R}_\mu \,,
\notag\\
D_\mu \xi_L & = \partial_\mu \xi_L - i V_\mu \xi_L + i \xi_L {\mathcal L}_\mu \,,
\label{eq:Dxi}
\end{align}
where ${\mathcal R}_\mu$ and ${\mathcal L}_\mu$ are the external gauge fields for $\mbox{SU(2)}_R \times \mbox{SU(2)}_L $ symmetry.
From the above covariant derivatives the two HLS one-forms follow,
\begin{align}
\hat\alpha_{\perp \mu} & = \frac{1}{2i} \left[ D_\mu \xi_R \cdot \xi_R^\dag - D_\mu \xi_L \cdot \xi_L^\dag \right],
\notag\\
\hat\alpha_{\parallel \mu} & = \frac{1}{2i} \left[ D_\mu \xi_R \cdot \xi_R^\dag + D_\mu \xi_L \cdot \xi_L^\dag \right].
\label{eq:oneforms}
\end{align}
Substituting Eq.~\eqref{eq:Mdecomp} into Eq.~\eqref{eq:LS}, the scalar sector takes the equivalent form
\begin{align}
{\mathcal L}_{S} = {} & \frac{1}{2} \partial_\mu \sigma\, \partial^\mu \sigma
+ \sigma^2\, \tr\left[ \hat\alpha_\perp^\mu \hat\alpha_{\perp \mu} \right]
\notag\\
& + \frac{1}{2} \partial_\mu\chi\partial^\mu\chi - V_0\left( \sigma, \chi \right),
\label{eq:LS2}
\end{align}
where $V_0(\sigma,\chi) \equiv V(M,\chi)\big|_{M = \sigma}$ follows from Eq.~\eqref{eq:potential} with $\tr[M^\dag M] = 2\sigma^2$ and $\tr[(M^\dag M)^2] = 2\sigma^4$.

The scalar ($\chi$--$\sigma$) mass matrix ${\mathcal M}^2_{ab} = \partial^2 V_0/\partial\phi_a \partial\phi_b |_{\rm vac}$, with $\phi_{a} \in \{\chi,\sigma\}$, follows from the curvature of $V_0$ of Eq.~\eqref{eq:LS2} at the vacuum point.
Diagonalizing this matrix yields a light and a heavy scalar state, with masses denoted by $m_\sigma$ and $m_\chi$, respectively.
The scalar mixing angle is defined by $\tan 2\theta_S = 2 {\mathcal M}^2_{\chi\sigma} / ({\mathcal M}^2_{\sigma\sigma} - {\mathcal M}^2_{\chi\chi})$, a vacuum quantity.

The vector-meson Lagrangian, with the mass terms dressed by the scale compensator, reads
\begin{align}
{\mathcal L}_{V} = {} & \frac{m_\rho^2}{g_\rho^2}\, \tilde\chi^2\, \tr\left[ \hat\alpha_{\parallel\mu} \hat\alpha_\parallel^\mu \right]
\notag\\
& + \left( \frac{m_\omega^2}{2 g_\omega^2} - \frac{m_\rho^2}{2 g_\rho^2} \right) \tilde\chi^2\, \tr\left[ \hat\alpha_{\parallel\mu} \right] \tr\left[ \hat\alpha_\parallel^\mu \right]
\notag\\
& - \frac{1}{2 g_\rho^2} \tr\left[ V_{\mu\nu} V^{\mu\nu} \right]
\notag\\
& - \left( \frac{1}{4 g_\omega^2} - \frac{1}{4 g_\rho^2} \right) \tr\left[ V_{\mu\nu} \right] \tr\left[ V^{\mu\nu} \right],
\label{eq:LV}
\end{align}
where $m_\rho$ and $m_\omega$ denote Lagrangian parameters. The last two terms are the HLS kinetic terms which do not contribute to static homogeneous matter. The scalar and vector sectors are coupled further by
\begin{equation}
{\mathcal L}_{SV} = \frac{C_{Vs}}{2}\, \sigma^2\, \tr\left[ \hat\alpha_{\parallel\mu} \hat\alpha_\parallel^\mu \right],
\label{eq:LSV}
\end{equation}
with $C_{Vs}$ a dimensionless coupling.
Then physical mass of vector mesons are expressed as
\begin{equation}
m_{V,\rm phys}^2 = m_V^2 + \tfrac{1}{2} C_{Vs}\, g_V^2 f_\pi^2\,, \qquad V = \rho, \omega\,,
\label{eq:mvphys}
\end{equation}
and thus differ in general from the bare parameters appearing in Eq.~\eqref{eq:LV}. Here and in the in-medium vector-mass relations of Sec.~\ref{sec:medmasses}, $g_V$ stands for $g_\rho$ or $g_\omega$.
\paragraph*{Nucleon sector.}
The nucleons are described by the mirror-assignment doublet $(\psi_1,\psi_2)$, whose chirality components $\psi_{\alpha l} = P_L \psi_\alpha$ and $\psi_{\alpha r} = P_R \psi_\alpha$, with $P_{R,L} = (1 \pm \gamma_5)/2$, transform as
\begin{align}
\psi_{1r} \to g_R \psi_{1r}\,, & \qquad \psi_{1l} \to g_L \psi_{1l}\,,
\notag\\
\psi_{2r} \to g_L \psi_{2r}\,, & \qquad \psi_{2l} \to g_R \psi_{2l}\,,
\label{eq:mirror}
\end{align}
where the interchanged chiral assignment of $\psi_2$ allows for a chiral invariant mass term. The nucleon part of the Lagrangian is
\begin{widetext}
\begin{align}
{\mathcal L}_{N} = {} & \sum_{\alpha=1,2} \sum_{s=l,r} \bar\psi_{\alpha s}\, i\gamma^\mu \partial_\mu \psi_{\alpha s}
- {m}_0\,\tilde\chi \left[\bar\psi_{1l}\psi_{2r}-\bar\psi_{1r}\psi_{2l}
-\bar\psi_{2l}\psi_{1r}+\bar\psi_{2r}\psi_{1l}\right]
\notag\\
& {}-g_1\left[\bar\psi_{1r} M^\dag \psi_{1l}+\bar\psi_{1l} M\psi_{1r}\right]
-g_2 \left[\bar\psi_{2r} M \psi_{2l} +\bar\psi_{2l} M^\dag \psi_{2r}\right]
\notag\\
& {}+ a_{\rho NN}\, F_{\beta'}(\tilde\chi) \Big[\bar\psi_{1l} \gamma^\mu
\big(\xi_L^\dag\hat\alpha_{\parallel\mu}\xi_L\big)\psi_{1l}
+\bar\psi_{1r} \gamma^\mu \big(\xi_R^\dag\hat\alpha_{\parallel\mu}\xi_R\big) \psi_{1r}
\notag\\
& \hspace{7.2em} {}
+\bar\psi_{2l} \gamma^\mu \big(\xi_R^\dag\hat\alpha_{\parallel\mu}\xi_R\big)\psi_{2l}
+\bar\psi_{2r} \gamma^\mu \big(\xi_L^\dag\hat\alpha_{\parallel\mu}\xi_L\big) \psi_{2r}\Big]
\notag\\
& {}+ a_{0NN}\, G_{\beta'}(\tilde\chi)\, \tr\left[ \hat\alpha_{\parallel\mu} \right]
\sum_{\alpha=1,2}\left( \bar\psi_{\alpha l} \gamma^\mu \psi_{\alpha l} + \bar\psi_{\alpha r} \gamma^\mu \psi_{\alpha r}\right),
\label{eq:LN}
\end{align}
\end{widetext}
where $m_0$ is the chiral invariant mass, $g_1$ and $g_2$ are the two Yukawa couplings, and $a_{\rho NN}$ and $a_{0NN}$ are the dimensionless isovector and isoscalar vector--nucleon couplings. The chiral invariant mass term is dressed by a single power of $\tilde\chi$, whereas the vector--nucleon couplings carry factors
\begin{align}
F_{\beta'}(\tilde\chi) & = C_{V_\rho} + \left(1-C_{V_\rho}\right) \tilde\chi^{\beta'}\,,
\notag\\
G_{\beta'}(\tilde\chi) & = C_{V_0} + \left(1-C_{V_0}\right) \tilde\chi^{\beta'}\,,
\label{eq:FG}
\end{align}
which reduce to unity in vacuum and interpolate between scale-invariant couplings ($C_V = 1$) and fully compensator-dressed ones ($C_V = 0$), as suggested by skyrmion-crystal and chiral-scale effective field theory analyses~\cite{Park:2008zg} and are crucial for understanding the quenched $g_A$ problem in nuclear physics~\cite{Ma:2020tsj}. We use the same $\beta'$ in these coupling functions and in the potential term of Eq.~\eqref{eq:potential}, thereby adopting a simplifying one-parameter assumption.

\paragraph*{Mean fields and energy density.}
First of all, we would like to stress that we adopt the canonical formalism to construct the energy density in the RMF approximation in this paper. For readers who are more familiar with the grand-canonical formalism, we summarize the relation between the two formalisms in Appendix~\ref{app:gc}.
At zero temperature we adopt the RMF approximation in the unitary gauge of the HLS and set the $\pi$ mean field to zero, 
$\xi_L = \xi_R = 1$, retaining the static mean fields $\phi = \{\tilde\chi, \sigma, \omega, \rho\}$ with $\omega \equiv \langle\omega_0\rangle$ and
$\rho \equiv \langle\rho_0^3\rangle$,
and replacing the nucleon bilinears by their expectation values, the vector and scalar densities of each species.\footnote{In this work, the baryon number densities are written as $n_i$ whereas $\rho$ stands for the rho meson mean field.}
After taking mean field $\hat\alpha_{\perp\mu} = 0$ and
$\hat\alpha_{\parallel\mu} = - V_\mu$,
so that the one-forms of Eq.~\eqref{eq:oneforms} reduce to
\begin{align}
\tr\left[ \hat\alpha_{\parallel\mu} \right] & = - g_\omega\, \omega\, \delta_{\mu 0}\,,
\notag\\
\hat\alpha_{\parallel\mu} - \tfrac{1}{2} \tr\left[ \hat\alpha_{\parallel\mu} \right] & = - g_\rho\, \rho\, \delta_{\mu 0}\, T_3\,,
\notag\\
\tr\left[ \hat\alpha_{\parallel\mu} \hat\alpha_\parallel^\mu \right] & = \tfrac{1}{2} \left( g_\rho^2 \rho^2 + g_\omega^2 \omega^2 \right),
\label{eq:alphaMF}
\end{align}
and hence ${\mathcal L}_{V} + {\mathcal L}_{SV}$ to 
\be
{\mathcal L}_{V} + {\mathcal L}_{SV} & = & \tfrac{1}{2} m_\omega^2 \tilde\chi^2 \omega^2 + \tfrac{1}{2} m_\rho^2 \tilde\chi^2 \rho^2 \nonumber\\
& &{} + \frac{C_{Vs}}{4}\,\sigma^2 (g_\rho^2 \rho^2 + g_\omega^2 \omega^2).
\ee
The nucleon species are labeled by parity and isospin, $i = (\pm; p, n)$, with $\tau_3 = +1$ for the proton and $\tau_3 = -1$ for the neutron, and both parity partners acquire the common vector shift
\begin{align}
\Sigma_{\tau_3}(\phi) = {} & \left[ \tfrac{1}{2}\, a_{\rho NN} F_{\beta'}(\tilde\chi) + a_{0NN}\, G_{\beta'}(\tilde\chi) \right] g_\omega\, \omega
\notag\\
& + \frac{\tau_3}{2}\, a_{\rho NN} F_{\beta'}(\tilde\chi)\, g_\rho\, \rho\,.
\label{eq:Sigma}
\end{align}
We write $\mu_i^\ast$ for the Fermi energy of species $i$ at the mean-field solution.

In the parity-doublet basis the nucleon mass matrix is isospin diagonal and given by
\begin{equation}
{\mathcal M}_N(\sigma,\tilde\chi) = \begin{pmatrix} g_1 \sigma & m_0 \tilde\chi \\ m_0 \tilde\chi & - g_2 \sigma \end{pmatrix},
\label{eq:massmatrix}
\end{equation}
obtained from Eq.~\eqref{eq:LN} after the standard redefinition $\psi_2 \to \gamma_5\psi_2$, which removes $\gamma_5$ from the {chiral invariant} mass term and flips the sign of the $g_2$ diagonal entry. The physical baryons are the eigenstates of ${\mathcal M}_N$ at each density with in-medium physical masses
\begin{equation}
m_\pm^\ast(\sigma,\tilde\chi)
= \frac{1}{2} \left[ \sqrt{ \left( g_1 + g_2 \right)^2 \sigma^2 + 4 m_0^2 \tilde\chi^2 } \mp \left( g_2 - g_1 \right) \sigma \right].
\label{eq:rotated}
\end{equation}

At fixed specie densities the energy density of the system is
\begin{equation}
\varepsilon\left( \{n_i\}; \phi \right) = \sum_{i} \left[ \varepsilon_{\rm kin}\!\left(n_i, m_i^\ast\right) + \Sigma_{\tau_3(i)}(\phi)\, n_i \right] + \overline V\!\left( \phi \right),
\label{eq:eps}
\end{equation}
where $\varepsilon_{\rm kin}$ is the zero-temperature free-Fermi-gas energy density with spin degeneracy two, the second term is the vector shift of Eq.~\eqref{eq:Sigma} carried by each occupied state, and the mesonic potential, measured from its vacuum value, reads
\begin{widetext}
\begin{align}
\overline V\!\left( \phi \right) = {} & -\frac{1}{2}\bar\mu^2 \left( \tilde\chi^2\sigma^2 - f_\pi^2 \right)
+ \frac{\lambda}{4} \left( \sigma^4 - f_\pi^4 \right)
+ \frac{h_4}{4}\left( \tilde\chi^4 - 1 \right)
+ \frac{h_5}{4+\beta'}\left( \tilde\chi^{4+\beta'} - 1 \right)
\notag\\
& - \frac{1}{2} m_\omega^2 \tilde\chi^2 \omega^2 - \frac{1}{2} m_\rho^2 \tilde\chi^2 \rho^2
- \frac{C_{Vs}}{4}\, \sigma^2 \left( g_\rho^2 \rho^2 + g_\omega^2 \omega^2 \right).
\label{eq:Vbar}
\end{align}
\end{widetext}

\paragraph*{Stationary conditions.}
The mean fields follow from the Euler--Lagrange equations of the four meson fields, whose nucleon sources are the densities. {The EOMs are from the stationary conditions of the energy density at fixed $\{n_i\}$ (Appendix~\ref{app:gc})}. The {nucleon densities} are the independent variables: each $n_i$ fixes a Fermi momentum $k_{F,i} = (3\pi^2 n_i)^{1/3}$ and hence $\mu_i^\ast=\sqrt{k_{F,i}^2 + m_i^{\ast 2}}$, and no chemical potential enters as an input. The scalar densities are expressed as
\begin{equation}
n_{s,i} = \frac{\partial\varepsilon_{\rm kin}}{\partial m_i^\ast}
= \frac{(m_i^\ast)^3}{2\pi^2}\left[ x_i\sqrt{1+x_i^2} - \mathrm{arcsinh}\, x_i \right],
\label{eq:rhos}
\end{equation}
with $x_i = k_{F,i}/m_i^\ast$. {Only the positive-parity states are populated in this work---the negative-parity partner {has}
no Fermi sea at any density}---so, writing $n_{s,i}$ for the scalar density of the positive-parity state of isospin $i \in \{p,n\}$, the stationary conditions actually solved read
\begin{widetext}
\begin{align}
\tilde\chi: \quad &
\frac{2 m_0^2 \tilde\chi}{W} \sum_{i=p,n} n_{s,i}
\notag\\
& + \beta' \tilde\chi^{\beta'-1} \sum_i n_i \left[ \tfrac{1}{2}\left(1-C_{V_\rho}\right) a_{\rho NN} \left( g_\omega \omega + \tau_3 g_\rho \rho\right)
+ \left(1-C_{V_0}\right) a_{0NN}\, g_\omega \omega \right]
\notag\\
& - \bar\mu^2 \tilde\chi \sigma^2 + h_4 \tilde\chi^3 + h_5 \tilde\chi^{3+\beta'}
- \left( m_\omega^2 \omega^2 + m_\rho^2 \rho^2 \right) \tilde\chi = 0\,,
\label{eq:statchi}\\[4pt]
\sigma: \quad &
\frac{1}{2}\left[ \frac{\left(g_1+g_2\right)^2 \sigma}{W} - \left(g_2-g_1\right) \right] \sum_{i=p,n} n_{s,i}
- \bar\mu^2 \tilde\chi^2 \sigma + \lambda \sigma^3
- \frac{C_{Vs}}{2}\, \sigma \left( g_\rho^2 \rho^2 + g_\omega^2 \omega^2 \right) = 0\,,
\label{eq:statsig}\\[4pt]
\omega: \quad &
\left[ m_\omega^2 \tilde\chi^2 + \frac{C_{Vs}}{2}\, g_\omega^2 \sigma^2 \right] \omega
= \left[ \tfrac{1}{2}\, a_{\rho NN} F_{\beta'}(\tilde\chi) + a_{0NN}\, G_{\beta'}(\tilde\chi) \right] g_\omega \sum_i n_i\,,
\label{eq:statomg}\\[4pt]
\rho: \quad &
\left[ m_\rho^2 \tilde\chi^2 + \frac{C_{Vs}}{2}\, g_\rho^2 \sigma^2 \right] \rho
= \frac{1}{2}\, a_{\rho NN} F_{\beta'}(\tilde\chi)\, g_\rho \sum_i \tau_{3}(i)\, n_i\,,
\label{eq:statrho}
\end{align}
\end{widetext}
where $W \equiv \sqrt{(g_1+g_2)^2\sigma^2 + 4 m_0^2 \tilde\chi^2}$ is the square root part
of Eq.~\eqref{eq:rotated}.
The coefficients in front of the scalar source $n_{s,i}$
are precisely $\partial m_+^\ast/\partial\tilde\chi$ and $\partial m_+^\ast/\partial\sigma$, the derivatives of the eigenvalue mass, and they have density dependence
through the condensates.
Because the mass operator is diagonal in the in-medium eigenbasis, no off-diagonal condensate arises, and Eqs.~\eqref{eq:statchi}--\eqref{eq:statrho} are exactly the stationarity conditions $\partial\varepsilon/\partial\phi_a = 0$ of the energy density of Eq.~\eqref{eq:eps} at fixed species densities.
Solving Eqs.~\eqref{eq:statchi}--\eqref{eq:statrho} at given $\{n_i\}$ determines the four mean fields and hence, through Eq.~\eqref{eq:eps}, the energy density.
The chemical potentials are defined as $\mu_i = \partial\varepsilon/\partial n_i$. Because the mean fields satisfy $\partial\varepsilon/\partial\phi_a = 0$, their implicit dependence on the densities does not contribute to this derivative, and differentiating Eq.~\eqref{eq:eps} at fixed $\phi$ gives
\be
\mu_i & = & \frac{\partial\varepsilon_{\rm kin}}{\partial n_i} + \Sigma_{\tau_3(i)}
= \sqrt{k_{F,i}^2 + m_i^{\ast 2}} + \Sigma_{\tau_3(i)}\nonumber\\
& = & \mu_i^\ast + \Sigma_{\tau_3(i)}\,,
\label{eq:mu}
\ee
the Fermi energy shifted by the vector self-energy of Eq.~\eqref{eq:Sigma}, which is the form quoted below that equation.

In practice $\mu_i$ is evaluated numerically by differentiating the tabulated energy per baryon (Sec.~\ref{sec:fit}), and the pressure follows from the Euler relation $P = \sum_i n_i \mu_i - \varepsilon$.
Appendix~\ref{app:gc} records the grand-canonical formulation of this model and shows that it yields the same mean-field equations, so that results obtained either way are consistent.

\subsection{Mean-field order parameters for chiral and scale symmetries}
\label{sec:order}

The energy density and the stationary equations of the mean fields were given in Eqs.~\eqref{eq:eps}--\eqref{eq:statrho}. Solving them at each baryon density $n$ for a prescribed composition---here symmetric NM, $n_p = n_n = n/2$---defines the mean-field profiles $\langle\sigma\rangle^\ast$ and $\langle\chi\rangle^\ast$. Within this treatment we study the two normalized scalar mean fields,
\begin{equation}
\Phi_\pi(n) \equiv \frac{\langle\sigma\rangle^\ast}{f_\pi}\,, \qquad
\Phi_\chi(n) \equiv \frac{\langle\chi\rangle^\ast}{f_\chi}\,.
\label{eq:orderparams}
\end{equation}
At tree level the coefficient of the pion kinetic term in Eq.~\eqref{eq:LS2} is $\sigma^2$, so $\Phi_\pi$ is the natural in-medium generalization of the ratio that defines $f_\pi$ in vacuum, and $\Phi_\chi$ is the corresponding quantity for the conformal compensator. Neither should be read as a decay constant in the sense of a current-correlator pole residue: a medium breaks Lorentz invariance, so the physical pion possesses distinct temporal and spatial constants that this truncation does not separate, and the strong $\chi$--$\sigma$ mixing (Sec.~\ref{sec:lagrangian}) means the scalar poles carry mixtures of $\langle\chi\rangle^\ast$ and $\langle\sigma\rangle^\ast$ rather than either alone.
Note that $\Phi_\pi$ defined here is not the quantity extracted from deeply bound pionic atoms~\cite{Suzuki:2002ae,Kienle:2004hq,Nishi:2022vib}.
We take $\Phi_\pi$ and $\Phi_\chi$ as chiral and scale order parameters, and compare them with empirical decay constants only indicatively.

With these distinctions, the density evolution of $\Phi_\pi$ and $\Phi_\chi$ is the cleanest
diagnostic of the two symmetry breakings in this model, and Sec.~\ref{sec:atdensity} sets it against the frozen-dilaton construction.

\section{{Fitting and results}}
\label{sec:fit}

We first explain how the parameters are determined, then present the fitted results in comparison with the frozen-dilaton $\sigma^6$-PDM~\cite{Minamikawa:2020jfj}.
Finally we discuss the properties of in-medium hadrons.

\subsection{{Fitting procedure}}
\label{sec:fitting}

\subsubsection{Vacuum inputs and free parameters}

The pion decay constant is fixed at $f_\pi = 92.4$~MeV and $n_0 = 0.16~\mathrm{fm}^{-3}$ sets the density scale\footnote{In this paper we use $n_0 = 0.16~\mathrm{fm}^{-3}$ as the normalization scale, and $n_s$ refers to saturation density for each model.}. The saturation density of the model itself comes out at $n_s = 0.164$~fm$^{-3}$ ($1.02\,n_0$). The vacuum stationary conditions~\eqref{eq:vacstat} eliminate $\bar\mu^2$ and the quartic coupling $\lambda$. 
All remaining parameters---$m_0$, $f_\chi$, $\beta'$, the potential couplings, the parameters $m_\rho$ and $m_\omega$ in Eq.~\eqref{eq:LV} and couplings, $g_{1,2}$, and the vector--nucleon couplings $a_{\rho NN}$, $a_{0NN}$, $C_{V_\rho}$, $C_{V_0}$, $C_{Vs}$---are determined numerically by the global fit, with the vacuum hadron masses ($m_N$, $m_{N^\ast}$, $m_\rho$, $m_\omega$, and the scalar eigenvalues) entering as $\chi^2$ targets rather than exact constraints (Table~\ref{tab:targets}). In addition, the prior constraints in Tab.~\ref{tab:bounds} are applied before fitting.

\subsubsection{Fitting targets and objective}

The objective {function} used for the fit is the sum of a NM and astrophysical contribution,
\begin{equation}
\chi^2_{\rm tot} = \chi^2_{\rm NM} + \chi^2_{\rm MR}\,.
\label{eq:chi2tot}
\end{equation}

The NM contribution collects the empirical targets of Table~\ref{tab:targets}.
The first term of Eq.~\eqref{eq:chi2tot} reads
\begin{equation}
\chi^2_{\rm NM} = \sum_i \left[\frac{O_i - \bar O_i}{\sigma_i}\right]^2
+ \left[\frac{E_d}{\sigma_d}\right]^2 ,
\label{eq:chi2nm}
\end{equation}
where the $O_i$ are the model outputs and $(\bar O_i, \sigma_i)$ denote the corresponding target values and widths listed in Table~\ref{tab:targets}.
The last term favors saturation at the reference density: $E_d \equiv d(E/A)/dn|_{n_0}$, equivalent to $P(n_0) = n_0^2 E_d$, is scored with the width $\sigma_d \simeq 3.8$~MeV\,fm$^3$ ($0.5\times10^{-6}$~MeV$^{-2}$ in the natural units), which is a soft condition. The best fit has $d(E/A)/dn|_{n_0} = -4.1$~MeV\,fm$^3$ [$P(n_0) = -0.11$~MeV\,fm$^{-3}$], and its energy minimum lies at $n_s = 0.164$~fm$^{-3}$. 

The astrophysical contribution $\chi^2_{\rm MR}$ is a sum over the three NICER pulsars of the squared distance from the adopted posterior summary to the closest point of the model mass--radius curve,
\begin{equation}
\chi^2_{\rm MR} = \sum_{p} \min_{(M,R)\,\in\,\rm curve}
\left\{ \left[\frac{M - M_p}{\sigma_M^{\,p}}\right]^2
+ \left[\frac{R - R_p}{\sigma_R^{\,p\mp}}\right]^2 \right\},
\label{eq:chi2mr}
\end{equation}
with the asymmetric radius width $\sigma_R^{\,p\mp}$ selected by the sign of $R - R_p$ (the asymmetric mass width of J0030 is symmetrized to its mean). These are credible-region distance measures, not full likelihood evaluations, and the adopted $(M_p, R_p, \sigma)$ are those of Table~\ref{tab:mrtargets}.
Both target sets are summarized in Tables~\ref{tab:targets} and~\ref{tab:mrtargets} together with their references.

The GW170817 constraint enters nowhere in Eq.~\eqref{eq:chi2tot}: it is checked a posteriori through $\tilde\Lambda$. The parameters are optimized in a two-phase search---a differential-evolution global exploration with local searching, followed by bounded Nelder--Mead refinement and a re-ranking of the surviving candidates by the exact Tolman--Oppenheimer--Volkoff (TOV) $\chi^2_{\rm MR}$---whose settings are collected in Appendix~\ref{app:repro}.

The hadronic EoS is obtained from the energy per baryon tabulated on a two-dimensional grid in the neutron and proton densities and interpolated by a quintic spline, following the conventions of Ref.~\cite{Ma:2025llw}, from which the chemical potentials follow by differentiation; beta equilibrium with electrons is imposed as $\mu_n = \mu_p + \mu_e$. A rough estimate shows that muons lower the pressure by about $3\%$ at $2n_0$, where the hadronic EoS ends.
A BPS (outer) plus Negele--Vautherin (inner) crust is attached, and the TOV equation is integrated for each candidate; the tidal deformability is obtained in post-processing from the Hinderer metric-perturbation equation~\cite{Hinderer:2007mb,Postnikov:2010yn}.
The saturation-point characteristics ($K_0$, $E_{\rm sym}$, $L$, $J_0$) are finite-difference curvatures of the hadronic $E/A$ at $n_s$---the connection never enters them---with stencils given in Appendix~\ref{app:repro}.

\begin{table}[htbp]
\begin{center}
\caption{
    Empirical targets entering $\chi^2_{\rm NM}$ Eq.~\eqref{eq:chi2nm}, with the literature supporting each adopted range.
    Several entries are syntheses or conventions rather than single posteriors---notably $m_\rho$, which is the $\rho(770)$ pole mass. The references supporting each range are given in the last column. }
\label{tab:targets}
\begin{tabular}{lccl}
\hline\hline
quantity & target & width & Refs. \\
\hline
$B/A$ [MeV] & $-16.0$ & $0.3$ & \cite{Margueron:2017lup,Drischler:2024ebw} \\
$E_{\rm sym}$ [MeV] & $32$ & $2$ & \cite{Margueron:2017lup,Lattimer:2023rpe} \\
$K_0$ [MeV] & $240$ & $20$ & \cite{Garg:2018uam} \\
$J_0$ [MeV] & $-300$ & $400$ & \cite{Xie:2020tdo,Dong:2025okn} \\
$L$ [MeV] & $55$ & $15$ & \cite{Margueron:2017lup,Lattimer:2023rpe,Drischler:2024ebw} \\
$m_N$ [MeV] & $939$ & $1$ & \cite{ParticleDataGroup:2024cfk} \\
$m_{N^\ast}$ [MeV] & $1530$ & $15$ & \cite{ParticleDataGroup:2024cfk} \\
$m_\sigma$ [MeV] & $600$ & $200$ & \cite{ParticleDataGroup:2024cfk,Pelaez:2015qba} \\
$m_\chi$ [MeV] & $1500$ & $1000$ & \cite{Morningstar:1999rf,Athenodorou:2020ani} \\
$m_\rho$ [MeV] & $763$ & $2$ & \cite{ParticleDataGroup:2024cfk} \\
$m_\omega$ [MeV] & $783$ & $1$ & \cite{ParticleDataGroup:2024cfk} \\
\hline\hline
\end{tabular}
\end{center}
\end{table}

\begin{table}[htbp]
\begin{center}
\caption{
    Adopted NICER pulsars entering $\chi^2_{\rm MR}$ Eq.~\eqref{eq:chi2mr} with the analyses each drawn from.
    The radius widths are asymmetric, $\sigma_R^{+}$ applying above $R_p$ and $\sigma_R^{-}$ below. The centres and widths are adopted from the analyses listed below. }
\label{tab:mrtargets}
\setlength{\tabcolsep}{4pt}
\begin{tabular}{lcc}
\hline\hline
pulsar & $M_p$ [$M_\odot$] & $R_p$ [km] \\
\hline
J0437$-$4715 & $1.418 \pm 0.037$~\cite{Choudhury:2024xbk,Reardon:2024rdv} & $11.36^{+0.95}_{-0.63}$~\cite{Choudhury:2024xbk} \\
J0030$+$0451 & {$1.34^{+0.15}_{-0.16}$~\cite{Riley:2019yda}} & {$12.71^{+1.14}_{-1.19}$~\cite{Riley:2019yda}} \\
J0740$+$6620 & {$2.073 \pm 0.069$~\cite{Salmi:2024aum}} & {$12.49^{+1.28}_{-0.88}$~\cite{Salmi:2024aum}} \\
\hline\hline
\end{tabular}
\end{center}
\end{table}

\subsubsection{The hadron--quark {transition}}
\label{sec:bridgedef}

Above a few times $n_0$ we expect the hadronic description to break down, and we use quark matter instead, described here by a tabulated 2+1-flavor NJL EoS (scalar coupling $G$, isoscalar-vector coupling $G_V$, diquark coupling $H$) precomputed following the conventions of {Ref.~\cite{Minamikawa:2020jfj}}.
The dimensionless ratios {$(H, G_V) \equiv (H/G, G_V/G)$} select the table member used in the fit (Appendix~\ref{app:repro}). Ref.~\cite{Minamikawa:2020jfj} writes this isoscalar-vector coupling as $g_V$; we rename it $G_V$ to keep it distinct from the hidden-local-symmetry vector couplings $g_\rho$ and $g_\omega$ of Eq.~\eqref{eq:LV}.
The hadronic and quark EoSs are joined by a quintic Hermite interpolation of $\varepsilon(n_B)$ (Appendix~\ref{app:repro}) between the hadronic EoS at a lower matching density $n_L$ and the NJL EoS at an upper density $n_H$. We refer to this construction as the ``polytrope'' bridge. The interpolation is carried out in the $\varepsilon$--$n$ plane. $\mu_B$ and $P$ follow from it by differentiation and by $P = n_B\mu_B - \varepsilon$.

The NJL table spans $5$--$10\,n_0$, and the bridge touches it solely through the endpoint values $\varepsilon$, $\mu$, and $d\mu/dn$ at $n_H$.
At the other end we have verified that every retained sequence reaches its maximum mass at a central density well inside the tabulated range, so no fit reported in this paper evaluates the quark model outside $[5,10]\,n_0$.

The bridge parameters are part of the fit in a nested sense: for each hadronic candidate the four connection parameters---$(H, G_V)$ together with $(n_L, n_H)$---are inner-optimized to minimize $\chi^2_{\rm MR}$.
A candidate is also required to yield a causal, mechanically stable connected EoS: if the bridge fails the build-stage sound-speed gate, no hybrid star is built for it (Appendix~\ref{app:repro}) si that no retained fit fails this gate. The construction presumes quark--hadron continuity. For the physics of why continuity stiffens the EoS see Ref.~\cite{Kojo:2025vcq}, and for the related quarkyonic realization Ref.~\cite{McLerran:2018hbz}. A Maxwell-type first-order transition, explored for this model class in Refs.~\cite{Gao:2024lzu,Yuan:2025dft}, would be the natural second variant; we do not pursue it here: the hadronic sector is already too soft to carry the massive pulsars (Sec.~\ref{sec:bridges}), and a first-order transition, which makes the EoS even softer at the jump, is expected to lower the maximum mass further.

\subsection{The fit and what the dilaton changes}
\label{sec:bridges}

We now present the fit selected by Eq.~\eqref{eq:chi2tot}.
It is the outcome of the two-step search of Appendix~\ref{app:repro}, iterated until the objective plateaued. We label the resulting best fit \textbf{sPDM}, and its quark matter is drawn from the tabulated NJL family with its own inner-optimized $(H, G_V)$.

It should be emphasized that in the present framework the hadron--quark connection is required, not merely tolerated. The purely hadronic EoS is causal and mechanically stable throughout but soft: with the bridge switched off its sequence tops out at $1.54\,M_\odot$ (dotted red curve in Fig.~\ref{fig:mr}), far short of the massive pulsars.
The connected sequence enters the bridge window already at $M \simeq 0.91\,M_\odot$ (indicated by the filled circle on the sPDM sequence), so every star above about one solar mass carries a bridge-affected core, whereas the hadronic branches of the frozen-dilaton $\sigma^6$ sets
alone reach $1.64$--$2.30\,M_\odot$.

\subsubsection{Best-fit parameters, nuclear matter, and stars}
\label{sec:bridgestars}

\begin{table*}[htbp]
\begin{center}
\caption{
    Best fit results of sPDM compared with the frozen-dilaton $\sigma^6$ sets of Ref.~\cite{Gao:2024chh}. The results of the latter are evaluated here from their published parameters, with the crust and the hadron--quark connection of this work and their $(H, G_V)$, with $m_\pi = 138$~MeV and electrons only in $\beta$ equilibrium.
    Dashes mark quantities with no analogue in a frozen-dilaton construction.
    $m_0$ and $g_{1,2}$ are fitted in this work, whereas in Ref.~\cite{Gao:2024chh} they are inputs fixed before that fit,
    in a common convention ($g_2 > g_1$). $J_0$ is a fitted target for sPDM but a pure output for the PDM sets. $\tilde\Lambda(q{=}1)$ is the equal-mass deformability at the GW170817 chirp mass.
    The $\chi^2$ block applies the exact rule of Eq.~\eqref{eq:chi2mr} to the tabulated sequences throughout. The NM rows of every column are evaluated at that set's own saturation density ($n_s = 0.164$~fm$^{-3}$ for sPDM), and the $\chi^2$ block is rescored at \(n_s\), while the fit minimized Eq.~\eqref{eq:chi2tot} at $n_0$.
    The vector weights $w_\omega$ and $w_\rho$, fixed in vacuum, are defined in Eq.~\eqref{eq:mvmix}.
    It should be noted that their \(m_\rho\) is chosen to be \(776 \rm\ MeV\) as input.
    }
\label{tab:bridges}
\label{tab:sig6}
\setlength{\tabcolsep}{7pt}
\begin{tabular}{lcccc}
\hline\hline
 & this work & \multicolumn{3}{c}{$\sigma^6$, frozen dilaton~\cite{Gao:2024chh}} \\
 & & \multicolumn{3}{c}{PDM-$m_0$, $m_0 =$} \\
 & sPDM & $600$ & $700$ & $800$ \\
\hline
\multicolumn{5}{l}{fitted parameters}\\
$m_0$ [MeV] & $657$ & $600$ & $700$ & $800$ \\
$f_\chi$ [MeV] & $98$ & --- & --- & --- \\
$\beta'$ & $0.286$ & --- & --- & --- \\
$g_1$ & $8.11$ & $8.48$ & $7.81$ & $6.99$ \\
$g_2$ & $14.52$ & $14.93$ & $14.26$ & $13.44$ \\
$h_4$ [$10^9\,$MeV$^4$] & $-36.96$ & --- & --- & --- \\
$h_5$ [$10^9\,$MeV$^4$] & $+40.00$ & --- & --- & --- \\
\hline
\multicolumn{5}{l}{derived vacuum spectrum}\\
$m_{S,\rm light}$ ($m_\sigma$) [MeV] & $629$ & $414$ & $388$ & $332$ \\
$m_{S,\rm heavy}$ [MeV] & $1462$ & --- & --- & --- \\
\hline
\multicolumn{5}{l}{hadron--quark bridge}\\
NJL $(H, G_V)$ & $(1.35, 0.80)$ & $(1.55, 1.30)$ & $(1.60, 1.30)$ & $(1.55, 1.00)$ \\
bridge window & $(1.99 \to 5.18)\,n_0$ & $(2 \to 5)\,n_0$ & $(2 \to 5)\,n_0$ & $(2 \to 5)\,n_0$ \\
\hline
\multicolumn{5}{l}{nuclear matter}\\
$n_s$ [fm$^{-3}$] & $0.164$ & $0.159$ & $0.159$ & $0.161$ \\
$B/A$ [MeV] & $-15.9$ & $-15.9$ & $-15.9$ & $-16.2$ \\
$K_0$ [MeV] & $259$ & $238$ & $238$ & $242$ \\
$E_{\rm sym}$ [MeV] & $31.8$ & $26.6$ & $26.5$ & $26.6$ \\
$L$ [MeV] & $72.2$ & $54.4$ & $50.2$ & $47.9$ \\
$J_0$ [MeV] & $-1801$ & $+572$ & $-76$ & $-370$ \\
\hline
\multicolumn{5}{l}{neutron stars}\\
$M_{\rm max}$ [$M_\odot$] & $2.30$ & $2.38$ & $2.35$ & $2.19$ \\
$R_{1.4}$ [km] & $12.22$ & $12.75$ & $12.12$ & $11.67$ \\
$M_{\rm max}^{\rm had}$ [$M_\odot$] & $1.54$ & $2.30$ & $1.95$ & $1.64$ \\
$\Lambda_{1.4}$ & $387$ & $541$ & $429$ & $401$ \\
$\tilde\Lambda(q{=}1)$ & $463$ & $646$ & $496$ & $489$ \\
\hline
\multicolumn{5}{l}{in-medium structure}\\
$\Phi_\pi$ / $\Phi_\chi$ at saturation & $0.76 / 0.93$ & $0.61 / 1$ & $0.66 / 1$ & $0.66 / 1$ \\
$w_\omega$ / $w_\rho$ & $0.04 / 0.36$ & --- & --- & --- \\
\hline
\multicolumn{5}{l}{fit quality (full objective, exact mass--radius rule)}\\
$\chi^2_{\rm NM}$ & $16.57$ & $55.37$ & $51.63$ & $52.26$ \\
\quad $\chi^2_{\rm sat}$ (nuclear matter) & $16.54$ & $12.12$ & $8.12$ & $8.00$ \\
\qquad of which $J_0$ & $14.09$ & $4.75$ & $0.31$ & $0.03$ \\
\quad $\chi^2_{\rm vac}$ (vacuum spectrum) & $0.03$ & $43.25$ & $43.51$ & $44.26$ \\
\qquad of which $m_\rho$ & $0.00$ & $42.25$ & $42.25$ & $42.25$ \\
$\chi^2_{\rm MR}$ & $1.05$ & $2.26$ & $1.12$ & $2.37$ \\
$\chi^2_{\rm tot}$ & $17.62$ & $57.63$ & $52.75$ & $54.63$ \\
\hline\hline
\end{tabular}
\end{center}
\end{table*}

\begin{figure}[htbp]
\includegraphics[width=\columnwidth]{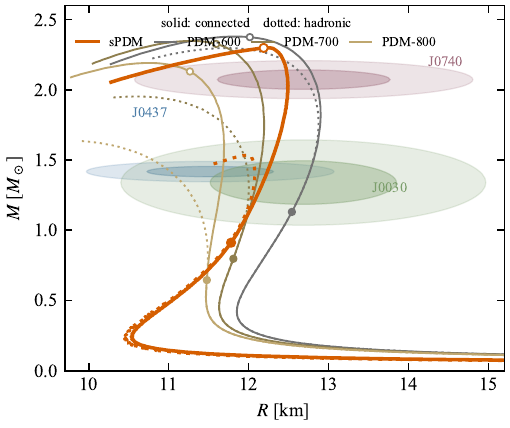}
\caption{
    Mass--radius relation for sPDM and the frozen-dilaton sets of Ref.~\cite{Gao:2024chh}.
    Dotted branches are the hadronic EoS with the bridge switched off.
    Circles mark the bridge window on the connected sequence, filled at $n_L$ and open at $n_H$.
    Ellipses approximate the $68\%$ and $95\%$ NICER credible regions adopted in the fit (Table~\ref{tab:mrtargets}). }
\label{fig:mr}
\end{figure}

From Table~\ref{tab:bridges}, one can see that every NM target is met within its window or close to its edge ($L$ sits slightly above its window); only the skewness falls far below the bulk of its constraint.
On the astrophysical side the fit reaches $\chi^2_{\rm MR} = $ $1.05$ under the exact closest-point rule, passing inside all three NICER credible regions (Fig.~\ref{fig:mr}).
The GW170817 deformabilities, which enter the objective nowhere, come out inside the observed intervals as well: $\Lambda_{1.4} = 387$ against $[70, 580]$ and $\tilde\Lambda(q{=}1) = 463$ against $[70, 720]$.

\begin{figure}[htbp]
\includegraphics[width=\columnwidth]{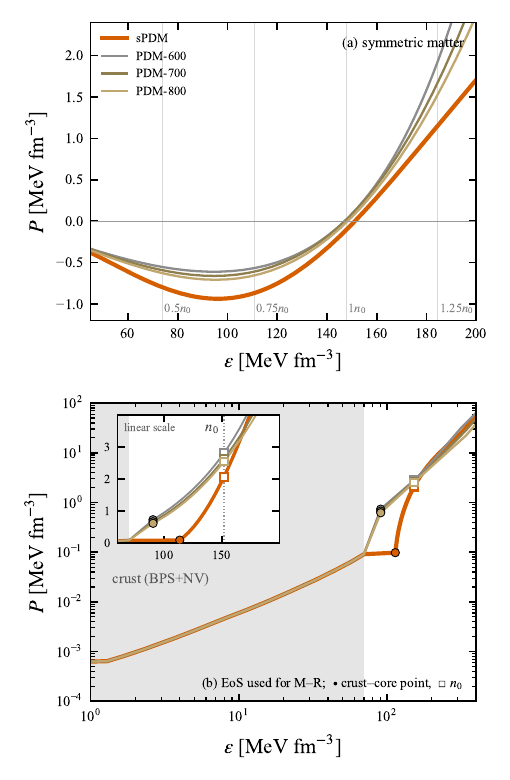}
\caption{(a) Pressure against energy density of symmetric NM for sPDM and the three frozen-dilaton sets. Vertical lines mark $\varepsilon$
at $0.5$, $0.75$, $1$ and $1.25\,n_0$. (b) The $\beta$-equilibrated EoS used for Fig.~\ref{fig:mr}. The shaded region is the crust; circles mark where the core joins the crust, and open squares mark $n_B = n_0$, where $\varepsilon \simeq 152$~MeV\,fm$^{-3}$ for all four sets. The inset shows this region on linear scales.}
\label{fig:soft}
\end{figure}

The low-mass part of the M--R relation is set by the hadronic EoS below about $1.5\,n_0$, and there the sPDM EoS is softer than those of the frozen-dilaton sets (Fig.~\ref{fig:soft}).
In symmetric matter the pressure of sPDM reaches approximately $-1.0$~MeV\,fm$^{-3}$ 
near $0.6\,n_0$, against $-0.6$ to $-0.7$~MeV\,fm$^{-3}$; in the $\beta$-equilibrated EoS its pressure at $n_0$ is around $2.1$~MeV\,fm$^{-3}$, against $2.6$--$2.8$~MeV\,fm$^{-3}$.
Because of this softness the core joins the crust at a lower pressure, $P_t = 0.1$~MeV\,fm$^{-3}$ at $n_t \approx 0.8\,n_0$ for sPDM, against $P_t = 0.6$--$0.7$~MeV\,fm$^{-3}$ at $n_t = 0.60\,n_0$ for the $\sigma^6$ sets.
Low-mass sPDM stars are therefore smaller, $R = 11.0$~km at $0.5\,M_\odot$ against $11.5$--$11.9$~km, while at $1\,M_\odot$ the radii are similar.

\subsubsection{Comparison with $\sigma^6$-PDM }
\label{sec:sigma6}

A closely related line of PDM studies---including prior works by some of the present authors and collaborators~\cite{Minamikawa:2020jfj,Gao:2022klm,Minamikawa:2023eky,Gao:2024chh,Gao:2024lzu,Kong:2025iqu}---is to stabilize the scalar potential with a $\sigma^6$ term instead of a dynamical dilaton, and connects to NJL quark matter at high density.
Structurally it is the frozen-dilaton limit of the present framework with an added sextic self-interaction, an $\omega$--$\rho$ mixing term, and an explicit chiral-symmetry-breaking term (finite $m_\pi$) that our chiral-limit construction lacks. The chiral invariant mass is a strict constant there, so the entire density evolution of the nucleon mass is carried by the $\sigma$ sector.

The parameter comparison is the sharpest part of Table~\ref{tab:bridges}. The fitted Yukawa couplings, $g_1 = 8.11$ and $g_2 = 14.52$, agree with those of PDM-600. And, the invariant mass $m_0=657$~MeV also falls between PDM-600 and PDM-700. However, the scalar mass is sensitive to the scalar potential. For example, with above mentioned values of $(m_0, g_1, g_2)$, the frozen-dilaton sets give light scalars of $m_{\sigma}^{\rm PDM}=332$--$414$~MeV while the dynamical dilaton gives $m_{\sigma}^{\rm sPDM}=629$~MeV. In frozen-dilaton PDM, the explicit-breaking term enters $m_\sigma^2$ only as the additive $m_\pi^2 \simeq 0.02$~GeV$^2$, an order of magnitude below the $(m_{\sigma}^{\rm{sPDM}})^2-(m_{\sigma}^{\rm{PDM}})^2\gtrsim 0.22$~GeV$^2$, so the chiral limit cannot by itself account for the difference. The two frameworks are thus distinguished by the scalar channel of the light meson spectrum.

The NM rows fill in the picture. Both frameworks sit on $n_s \simeq 0.16~\mathrm{fm}^{-3}$ and $B/A \simeq -16$~MeV by construction, and the incompressibilities are not far apart: $238$--$242$~MeV for frozen-dilaton PDM is centered on the target, while the result from sPDM $259$~MeV sits on the band's upper edge. The skewness row is the striking one. Evaluated with a standalone solver of that exact model, the frozen-dilaton family spans $J_0 = +572$ to $-1019$~MeV across $m_0 = 600$--$900$~MeV, its $m_0 = 700$ and $800$~MeV members falling inside our own target window $-300 \pm 400$~MeV---without any skewness information having entered the frozen-dilaton $\sigma^6$ fit. Our fit carries that target in its objective, pays $14.1$ units of $\chi^2$ for it, and still comes out at $-1801$~MeV. A frozen-dilaton $\sigma^6$ potential apparently produces a moderate skewness by construction; the dynamical dilaton does not. Whether that is intrinsic to the dilaton cannot be settled here. Including the explicit chiral-symmetry breaking, the $\sigma^6$ term and corrections to leading order scale symmetry used in this work might improve the situation.

One systematic statement can be made from the sPDM. Our reproduction of the published sixth-order sets confirms their symmetric-matter sector---$n_s$, $B/A$, and $K_0 \simeq 240$~MeV land on target for every $m_0$---and yields, via the curvature relation $m_\sigma^2 = 2(\lambda_4 - 2\lambda_6 f_\pi^2) f_\pi^2 + m_\pi^2$ applied to those tabulated couplings, a light scalar of $271$--$414$~MeV across the $m_0$ window. Their light scalars are thus locked to that range, whereas allowing a dynamical dilaton yields $629$~MeV at $K_0 = $ $259$~MeV.

\subsubsection{The two frameworks at density}
\label{sec:atdensity}

\begin{figure}[htbp]
\includegraphics[width=\columnwidth]{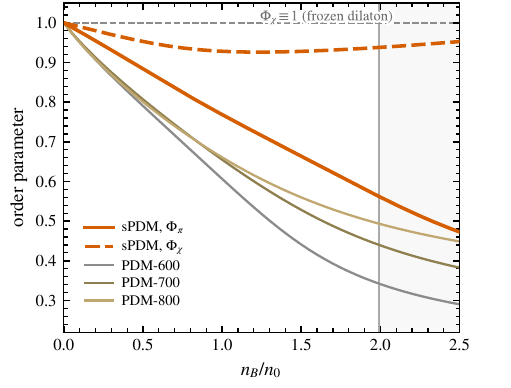}
\caption{Order parameters of Eq.~\eqref{eq:orderparams} in symmetric matter: the chiral order parameter $\Phi_\pi$ for sPDM (solid red) and for the three $\sigma^6$ sets, together with sPDM's scale order parameter $\Phi_\chi$ (dashed red). The frozen-dilaton construction fixes $\Phi_\chi \equiv 1$.}
\label{fig:order}
\end{figure}

The saturation points and the stars from both frameworks are similar and the density-resolved quantities are where the two frameworks separate. Figure~\ref{fig:order} shows the order parameters of Eq.~\eqref{eq:orderparams}. The $\sigma^6$ chiral condensates melt considerably faster than ours---$\Phi_\pi = 0.61$--$0.66$ at saturation and $0.34$--$0.49$ at $2\,n_0$, against $0.76$ and $0.56$ from sPDM---while their scale sector does not respond at all: $\Phi_\chi \equiv 1$ by construction. $\Phi_\chi$ from sPDM instead falls to a shallow minimum of $0.926$ near $1.24\,n_0$ and {rises} thereafter, reaching $0.94$ at the matching density. Chiral restoration is therefore well advanced within the hadronic window while the scale sector barely moves, and neither of our {order parameters} reaches zero before the quark sector takes over. That non-monotonic dilaton reappears in every quantity it feeds.

\begin{figure}[htbp]
\includegraphics[width=\columnwidth]{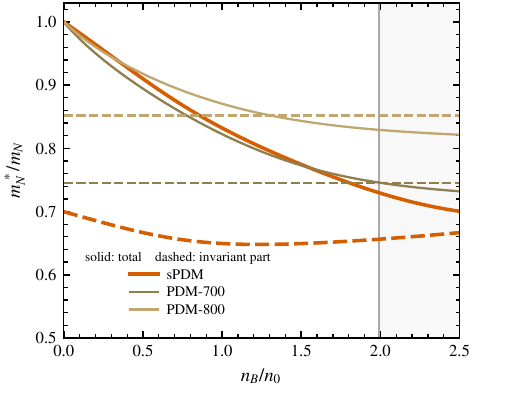}
\caption{Nucleon effective mass in symmetric matter, normalized to its vacuum value, for sPDM, PDM-700 and PDM-800. Solid curves are the total, dashed curves its invariant part---the constant $m_0$ in the frozen-dilaton model, the dynamical $m_0\tilde\chi$ here. The vertical line marks the two frameworks' bridge edges, which nearly coincide ($2\,n_0$ and $1.99\,n_0$).}
\label{fig:ghmed}
\end{figure}

Figure~\ref{fig:ghmed} turns the same comparison on the nucleon mass, each curve drawn with its {invariant part}---the piece that does not melt with the chiral condensate. The totals are strikingly similar: at saturation sPDM has lost $17\%$ of its mass, between PDM-700's $18\%$ and PDM-800's $13\%$, and the three curves cross repeatedly within the hadronic window. What distinguishes the frameworks is the composition of what remains, and how the invariant part gets there. In the frozen-dilaton sets it is a horizontal line by construction, and by the frozen-dilaton bridge edge at $2\,n_0$ both totals have converged onto the $m_0 = 700$ and $800$~MeV floors, PDM-800's dipping marginally below, so those models answer the question this paper poses---how much nucleon mass survives chiral restoration in medium.
In sPDM the invariant part is the dynamical curve $m_0\tilde\chi$, which decreases first with dilaton and then recovers: from vacuum to the matching density $n_L = 1.99\,n_0$ it falls from $657$ to $610$~MeV around saturation and returns to $616$~MeV, a net reduction of $6\%$, while the chirally generated remainder, \(m_N^*-m_0\tilde\chi\), collapses from $282$ to $69$~MeV, a reduction of $76\%$---the total going from $939$ to $685$~MeV. Chiral melting outpaces scale melting by an order of magnitude, so the invariant fraction of the nucleon mass rises from $0.70$ in vacuum to $0.90$ at the handover---delivered by the stationarity conditions rather than imposed. 

\subsection{In-medium structure of sPDM}
\label{sec:medium}

We turn now to the quantities the dilaton controls: the in-medium meson masses and the effective couplings.

\subsubsection{Scalar and vector masses}
\label{sec:medmasses}

The meson masses evolve with density, and the vector sector does so in a way that follows directly from the two order parameters of Fig.~\ref{fig:order}. The bracket multiplying $\omega$ in Eq.~\eqref{eq:statomg}, and its $\rho$ counterpart in Eq.~\eqref{eq:statrho}, is the in-medium squared vector mass $m_V^{\ast\,2}$: a dilaton contribution $m_V^2\tilde\chi^2$ plus a chiral contribution $\tfrac{1}{2} C_{Vs} g_V^2 \sigma^2$. In contrast to the nucleon case, where the two scales combine under a square root, they add here at the level of the squared mass, so the split is unambiguous. Dividing by the vacuum masses of Eq.~\eqref{eq:mvphys} normalizes it to
\begin{equation}
\frac{m_V^{\ast\,2}}{m_{V,\rm phys}^2}
= \left(1 - w_V\right)\Phi_\chi^2 + w_V\,\Phi_\pi^2\,,
\quad
w_V = \frac{C_{Vs}\, g_V^2 f_\pi^2}{2\,m_{V,\rm phys}^2}\,.
\label{eq:mvmix}
\end{equation}

\begin{figure}[htbp]
\includegraphics[width=\columnwidth]{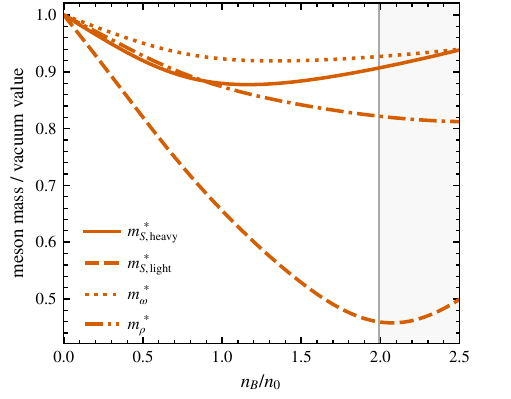}
\caption{
Four
in-medium meson masses of sPDM, each normalized to its vacuum value: the two eigenvalues of the $\chi$--$\sigma$ mass matrix and the two vector curvature masses of Eq.~\eqref{eq:mvmix}.}
\label{fig:medmasses}
\end{figure}

Figure~\ref{fig:medmasses} collects the four in-medium meson masses, each normalized to its vacuum value so that they share one axis; the ordering of the curves is a direct readout of how much chiral condensate each mass carries. The light scalar eigenvalue, predominantly $\sigma$-like at this set's mixing ($|\theta_S| = 25.2^\circ$), moves the most: from $629$~MeV in vacuum to $408$ at $n_s$ ($-35\%$) and $289$ at the matching density ($-54\%$), the strong chiral restoration of this fit dragging it down, though it stays comfortably massive throughout the window. The heavy eigenvalue, predominantly $\chi$-like, instead tracks the dilaton and moves much less, falling only to $0.88$ of its vacuum value: $1462$~MeV in vacuum, a minimum of $1283$~MeV near $1.16\,n_0$, and $1326$~MeV at matching, the recovery inherited from the dilaton itself. The two vector masses follow Eq.~\eqref{eq:mvmix}, which fixes where each of them falls: the $\omega$, with only $4\%$ of its squared mass in the chiral channel, falls by at most $8\%$. The $\rho$, carrying $36\%$ of its squared mass in the chiral channel, falls monotonically to $0.82$ of its vacuum value at matching.

The two vector mesons therefore behave oppositely from this fit: an in-medium $\omega$ mass would measure the dilaton almost alone, whereas an in-medium $\rho$ mass is the one that would betray this fit's chiral restoration. This behavior is a result of the fitted parameters, not of the framework. What Eq.~\eqref{eq:mvmix} establishes is only the form of the present model, so that measuring an in-medium vector mass constrains one particular linear combination of $\Phi_\chi^2$ and $\Phi_\pi^2$ whose coefficients must be reported for each fit.

\subsubsection{Effective couplings}
\label{sec:effcoup}

The couplings run with dnsity as well. In the scalar sector the density dependence is purely kinematic: the nucleon couples to the two condensates through the derivatives of Eq.~\eqref{eq:rotated} that already appear {as the coefficients multiplying the scalar densities $n_{s,i}$ in} Eqs.~\eqref{eq:statchi}--\eqref{eq:statsig}, which we write $g_{\sigma NN}^{\,\ast} \equiv \partial m_+^\ast/\partial\sigma$ and $g_{\chi NN}^{\,\ast} \equiv \partial m_+^\ast/\partial\chi = f_\chi^{-1}\partial m_+^\ast/\partial\tilde\chi$. They run only because the mass formula is nonlinear. For sPDM that nonlinearity is not a small correction: from vacuum to the matching density $g_{\sigma NN}^{\,\ast}$ falls from $6.38$ to $4.61$ ($-28\%$) and $g_{\chi NN}^{\,\ast}$ rises from $3.56$ to $4.85$ ($+36\%$)---as the condensate melts, the nucleon progressively decouples from the $\sigma$ direction and couples more strongly to the dilaton. The melting of the nucleon mass documented above is nonetheless carried chiefly by the movement of the condensates themselves.

The vector sector is different. There the couplings carry an explicit dilaton dependence through the scale-covariant functions of Eq.~\eqref{eq:FG}---$F_{\beta'}$, which multiplies the isovector coupling $a_{\rho NN}$, and $G_{\beta'}$, which multiplies the isoscalar $a_{0NN}$---both reducing to unity in vacuum. Reading the coefficients off Eq.~\eqref{eq:Sigma}, the physical vector--nucleon couplings are $g_{\rho NN}^{\,\ast} = a_{\rho NN}F_{\beta'}\,g_\rho/2$ and $g_{\omega NN}^{\,\ast} = [\,a_{\rho NN}F_{\beta'}/2 + a_{0NN}G_{\beta'}\,]\,g_\omega$. The prefactor $1-C_{V_i}$ is the amplifier, and because $\tilde\chi$ is depleted by only some $7\%$ it alone decides whether these functions stay near unity or are thrown far from it. Here the amplified channel is the isovector one: with $C_{V_\rho} = -21.1$ the prefactor is $22.1$, converting the modest dilaton dip into a factor-of-two suppression---$F_{\beta'}$ falls to $0.52$ at the dilaton minimum and, because the dilaton itself turns around, recovers to $0.60$ at matching, without changing sign. The isoscalar function is nearly inert by comparison ($C_{V_0} = +5.2$, prefactor $-4.2$): $G_{\beta'}$ rises gently to at most $1.09$. In the physical couplings $|g_{\rho NN}^{\,\ast}|$ follows $F_{\beta'}$, dropping from $6.66$ to $3.45$ and recovering to $3.98$, while $|g_{\omega NN}^{\,\ast}|$ moves oppositely, from $4.84$ to $6.47$ at $n_0$ before settling at $6.25$.

\begin{figure}[htbp]
\includegraphics[width=\columnwidth]{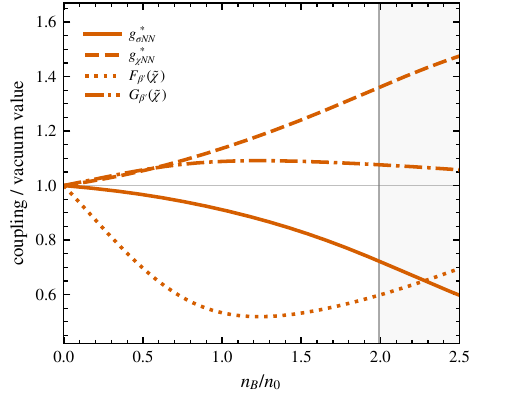}
\caption{Effective couplings of sPDM, normalized to their vacuum values: the scalar couplings $g_{\sigma NN}^{\,\ast}$ and $g_{\chi NN}^{\,\ast}$, and the scale-covariant vector functions of Eq.~\eqref{eq:FG}.}
\label{fig:vsplit}
\end{figure}

Setting number games aside, what survives is that reproducing NM in this framework prefers a vector coupling function with a strong scale dependence in at least one channel. That is the useful content of Fig.~\ref{fig:vsplit}: it identifies $C_{V_\rho}$ and $C_{V_0}$ as the parameters most in need of an independent handle, since each can turn a few-percent dilaton depletion into an order-unity change in a coupling.

\section{Summary and outlook}
\label{sec:summary}

We have constructed and fitted a scale-invariant PDM in which the effective chiral invariant mass of nucleon scales in medium as $m_{\rm inv}^\ast = m_0\tilde\chi$. The effective nucleon mass includes the contributions from both the spontaneous chiral symmetry breaking and the effective chiral invariant part $m_{\rm inv}^\ast$.
The fitting was carried out in the chiral limit against NM and selected NICER results, with the hadronic EoS joined to tabulated 2+1-flavor NJL quark matter by {a smooth polynomial interpolation of $\varepsilon(n_B)$, with the NJL EoS used only in its tabulated range $5$--$10\,n_0$}. The best fit yields $m_0 = 657$~MeV with $M_{\rm max} = 2.30\,M_\odot$ and $R_{1.4} = 12.22$~km, the mass--radius sequence passing through all three NICER credible regions, and predicted tidal deformabilities consistent with GW170817~\cite{LIGOScientific:2018hze,LIGOScientific:2018cki} which were not included in the fit. The one empirical target that remains clearly unmet is the skewness, $J_0 = $ $-1801$~MeV, against $-300 \pm 400$~MeV.
Whether a sharply narrowed skewness window could force the empirical range, and at what cost, is a question for the ensemble study proposed below.

Evaluating the frozen-dilaton $\sigma^6$-PDM sets of Ref.~\cite{Gao:2024chh} with the same $\chi^2$ shows the clear effect of dilaton. The $m_0 = 600$--$800$~MeV sets span $J_0 = +572$ to $-370$~MeV, the $m_0 = 700$ and $800$~MeV members sitting inside the window but
our fit gives $-1.8$~GeV. While the mass of the scalar meson is locked to $332$--$414$~MeV for the three sets in Table~\ref{tab:bridges}, the present model with dynamical dilaton (sPDM) gives $m_{\sigma}=629$~MeV, consistent with PDG~\cite{ParticleDataGroup:2024cfk}.

The last point concerns the in-medium behavior: The magnitude of chiral restoration is larger than that of scale restoration ($\Phi_\pi = 0.56$ against $\Phi_\chi = 0.94$ at the matching density~$n_L \approx 2\,n_0$), so 
the chiral invariant fraction of the nucleon mass $m_{\rm inv}^\ast/m_N^\ast$ rises from $0.70$ in vacuum to $0.90$ at $n_L$. The vector meson masses, whose squares are exact weighted means of the squared order parameters with weights fixed in vacuum, split into a nearly pure scale probe ($\omega$, falling by at most $8\%$) and a chirally mixed one ($\rho$, falling by $18\%$). In addition, the amplified isovector coupling function $F_{\beta'}$ is halved and partially recovers as the dilaton dips and rises.
The medium dependence of chiral invariant mass originates dynamically rather than an assumption.

Several directions follow directly. The most immediate direction is a quantified uncertainty statement. This paper presents a single best fit, and a systematic exploration of the parameter space around the best fit point---how tightly the data pin $m_0$, which of the in-medium statements of Sec.~\ref{sec:medium} are properties of the framework rather than of this fit, and whether the extreme skewness persists across an ensemble---is interesting. Promoting the GW170817 deformabilities from a posteriori check to fitted targets would be straightforward.

On the model side, restoring the explicit chiral-symmetry-breaking term, allowing a sextic scalar self-interaction alongside the dilaton, going beyond leading order scale symmetry, including muons in $\beta$ equilibrium, and extending to the SU(3) symmetry~\cite{Fraga:2023wtd,Gao:2026scv} are interesting.
Whether the hyperon onset thresholds found in the SU(3) analysis of Ref.~\cite{Gao:2026scv} carry over to the present two-flavor dilaton construction is not known, but at our $m_0 = 657$~MeV an onset inside the hadronic window remains plausible.
The finite-$m_\pi$ extension carries particular weight because it would make the $\sigma^6$ comparison fully like-for-like, testing directly whether the extreme output skewness is tied to the chiral limit. 

Furthermore, evaluating the normalized trace anomaly of Ref.~\cite{Fujimoto:2022ohj} along our connected EoSs, and placing the result against inferences of near-conformal matter in massive-star cores~\cite{Annala:2023cxo,Marczenko:2025hsh}---and against the newer quasi-universal stellar trace-anomaly profiles and heavy-ion flow constraints~\cite{Ren:2025xdx,Li:2026tjj}---would tie the dilaton sector of this model to a diagnostic {criterion} in the field~\cite{Ma:2026koy}.

\begin{acknowledgments}
The work of Y.~M. is supported by Jiangsu Funding Program for Excellent Postdoctoral Talent under Grant No.~2025ZB516. The work of B.~G. is supported in part by JSPS KAKENHI Grant No.~26K17147. The work of M.~H. is supported in part by JSPS KAKENHI Grant Nos.~23H05439 and 24K07045. The work of Y.-L.~M. is supported in part by the National Science Foundation of China (NSFC) under Grant No.~12675178 and~12547104, and Gusu Talent Innovation Program under Grant No.~ZXL2024363.
\end{acknowledgments}

\appendix
\counterwithin{table}{section}
\counterwithin{figure}{section}

\section{Numerical settings and reproducibility}
\label{app:repro}

The $\chi^2$ is defined by Eqs.~\eqref{eq:chi2tot}--\eqref{eq:chi2mr}. The target values and their references are listed in Tables~\ref{tab:targets} and~\ref{tab:mrtargets}. This appendix discusses the remaining numerical inputs. The $\chi^2$ term for the two-solar-mass constraint PSR J0740$+$6620 enters through Eq.~\eqref{eq:chi2mr}, the combined deviation from the measured $(M, R)$ to the nearest point of the mass--radius curve which is small only if the curve reaches about $2\,M_\odot$.

The complete code used for this work---the symbolic derivation of the mean-field equations, the solver, the construction of the EoS, the hadron--quark transition, and the solution of the TOV and tidal-deformability equations---is openly available at \url{https://github.com/AaronMahn/spdm-dilaton}.
The following items are about how we handle the numerical process.
\begin{itemize}
\item 
The saturation properties $K_0$, $L$ and $J_0$ are density derivatives of $E/A$ at the saturation density $n_s = 1.02\,n_0$, which we compute by finite differences. In symmetric matter the density step {size} is $\Delta n = 0.02\,n_0$ (a step {size} of $0.01\,n_0$ in each of $n_n$ and $n_p$). The third derivative in $J_0$ is computed from the four values of $E/A$ at $n = n_s \pm 0.02\,n_0$ and $n_s \pm 0.06\,n_0$. The symmetry energy is the curvature of $E/A$ in the asymmetry $\delta = (n_n - n_p)/n$, evaluated with $\delta = \pm 0.04$, and $L = 3 n_s\, dE_{\rm sym}/dn$ uses the same density step. The same formulas are used for all the fits in this paper.

\item 
The accuracy of these derivatives was checked for two quantities---$J_0$ and $K_0$---by reducing the step size to $\Delta n = 0.005\,n_0$ and extrapolating to zero step (Richardson extrapolation). The value of $J_0$ changes by less than $0.3\%$, and that of $K_0$ at the level of $0.01\%$. 

\item
For the hadronic EoS, $E/A$ is computed on an $81\times81$ grid in $(n_n/n_0, n_p/n_0)$ up to $8.5\,n_0$ and interpolated by fifth-order splines. During the fit a coarser $21\times21$ grid is used.

\item 
The TOV equations are solved by the fourth-order Runge--Kutta method with radial step $\Delta r = 0.05$~km, made smaller near the surface, for $400$ values of the central pressure. The $\beta$-equilibrium EoS is given as a table of $800$ points. At low density it is joined to the BPS (outer)~\cite{Baym:1971pw} and Negele--Vautherin (inner) crust~\cite{Negele:1971vb}, at the density of the model's Maxwell construction below $n_0$.
\end{itemize}

The $16$ fitted parameters and their ranges are listed in Table~\ref{tab:bounds}, {with $f_\pi$ and $n_0$ fixed}. We minimize $\chi^2_{\rm tot}$ in two steps. In the first step we search the whole parameter range of Table~\ref{tab:bounds} with the differential-evolution method. We start from $192$ parameter sets spread uniformly over the ranges and update them for at most $200$ iterations. The best set is then improved by a local gradient method (L-BFGS-B). In the second step the best sets are improved further by the Nelder--Mead method within the same ranges. In both steps the radii in $\chi^2_{\rm MR}$ are estimated by an approximate Lattimer--Prakash-type formula. At the end we recompute $\chi^2_{\rm MR}$ of the best sets by solving the TOV equations exactly, and we order the sets by this value.

In addition to the $\chi^2$, three conditions are imposed, and a parameter set that violates any of them is rejected.
(i) The physical vacuum masses, which also enter $\chi^2$ as targets, must lie in the ranges $740 \le m_{\rho,\rm phys} \le 790$~MeV, $760 \le m_{\omega,\rm phys} \le 806$~MeV, $800 \le m_N \le 1050$~MeV and $1200 \le m_{N^\ast} \le 2000$~MeV. Because of this condition the bare vector meson masses can be given the wide range $[0,1000]$~MeV in Table~\ref{tab:bounds}.
(ii) The hadronic $\beta$-equilibrium EoS must be causal and stable, $0 \le c_s^2 \le 1$ and $dP/dn > 0$. This is required for $1 \le n/n_0 \le 2.5$ when it is connected to quark matter, and for $1 \le n/n_0 \le 8$ when it is not. At higher density the hadronic EoS is replaced by the interpolation, so an instability there does not matter.
(iii) The full EoS, including the interpolation to quark matter, must also satisfy $0 \le c_s^2 \le 1$ and $dP/dn > 0$, otherwise no star is computed. None of the parameter sets reported in this paper violates this condition.

\begin{table}[htbp]
\begin{center}
\caption{Searching ranges of the fitted parameters.}
\label{tab:bounds}
\begin{tabular}{lcc}
\hline\hline
parameter & range & units \\
\hline
$a_{0NN}$, $a_{\rho NN}$ & $[-40, 40]$ & --- \\
$\beta'$ & $[0.1, 8.0]$ & --- \\
$C_{V_\rho}$ & $[-40, 40]$ & --- \\
$C_{V_0}$ & $[-60, 60]$ & --- \\
$C_{Vs}$ & $[-15, 15]$ & --- \\
$g_\rho$ & $[-30, 30]$ & --- \\
$g_\omega$ & $[-25, 25]$ & --- \\
$h_4$, $h_5$ & $[-10^{11}, 10^{11}]$ & MeV$^4$ \\
$g_1$, $g_2$ & $[-30, 30]$ & --- \\
$m_0$ & $[300, 1100]$ & MeV \\
$f_\chi$ & $[50, 600]$ & MeV \\
$m_\rho^{\rm bare}$, $m_\omega^{\rm bare}$ & $[0, 1000]$ & MeV \\
\hline\hline
\end{tabular}
\end{center}
\end{table}

Between $n_L$ and $n_H$ the energy density $\varepsilon(n_B)$ is written as a fifth-order polynomial in $t = (n_B-n_L)/(n_H-n_L)$. Its six coefficients are fixed by requiring that $\varepsilon$, $\mu_B = d\varepsilon/dn_B$ and $d\mu_B/dn_B$ agree with the hadronic EoS at $n_L$ and with the NJL EoS at $n_H$ (quintic Hermite interpolation). Since $P = n_B\mu_B - \varepsilon$, the pressure is then also continuous. (We use $t$ rather than $x$ because $x_i$ denotes the Fermi variable of Eq.~\eqref{eq:rhos}.)

The quark EoS is a precomputed table for 2+1-flavor NJL matter. It gives $\mu_B$, $P$, $\varepsilon$ and the $u$-quark constituent mass $M_u$ at $n_B/n_0 = 5$--$10$ in 51 points. The table covers the couplings $H = 1.05$--$1.75$ (step $0.025$) and $G_V = 0.1$--$1.7$ (step $0.05$), with 956 available pairs.
The interpolation uses this table only at $n_H$, and above $n_H$ the table itself is used.
Here $H$ and $G_V$ denote the ratios $H/G$ and $G_V/G$ to the scalar NJL coupling $G$, as in the quark--hadron crossover (QHC) construction~\cite{Baym:2017whm,Baym:2019iky} and in Ref.~\cite{Minamikawa:2020jfj}, where $G_V$ is written $g_V$. The best fit uses $(H, G_V) = (1.35, 0.80)$.

We now discuss the parameter counting. We fit sixteen hadronic parameters and the four parameters $(H, G_V, n_L, n_H)$ of the connection to quark matter to fifteen $\chi^2$ terms: the eleven entries of Table~\ref{tab:targets}, the saturation condition, and one term per pulsar. A simple count of the degrees of freedom therefore suggests overfitting, and $\chi^2_{\rm tot}$ cannot be used as a goodness-of-fit test in the usual sense. This count, however, underestimates the information in the data. A NM target constrains $E/A$ only near $n_0$. A mass--radius point, in contrast, is obtained by integrating the TOV equations from the center of the star to its surface, so it depends on $P(\varepsilon)$ over a wide density range. Each pulsar therefore constrains the shape of the EoS, not a single number. The same conclusion was reached in Refs.~\cite{Guo:2023mhf,Ma:2026kun,Ma:2026jff}, where a fit of seven parameters to six constraints gave an isolated minimum rather than a flat direction in parameter space. We find the same {conclusion} here: when $m_0$ is fixed at other values and the NM $\chi^2$ is minimized over the other parameters, $\chi^2$ increases steadily away from the best fit. A full scan of the parameter space is left to future work. We therefore use $\chi^2_{\rm tot}$ to rank the parameter sets compared in this paper, not as an absolute measure of fit quality, and we see no signal of overfitting.

\section{Relation to the grand-canonical formulation}
\label{app:gc}

{
In Sec.~\ref{sec:lagrangian} we solve the EOMs in the canonical formulation within the RMF approximation.}
Another treatment of PDMs at finite density is to use the grand canonical potential, with the chemical potentials introduced at the Lagrangian level as background values of external gauge fields.
In this appendix we summarize that formulation and show that it gives the same mean-field equations, so that results of the two formulations can be compared directly.

Chemical potentials are introduced as background values of external gauge fields: ${\mathcal L}_\mu = \sum_a {\mathcal L}_\mu^a T_a$ and ${\mathcal R}_\mu = \sum_a {\mathcal R}_\mu^a T_a$ for $\mathrm{SU}(2)_L \times \mathrm{SU}(2)_R$, and ${\mathcal V}^{(0)}_\mu$ for $\mathrm{U(1)}_B$. Coupling them minimally promotes the derivatives of Eqs.~\eqref{eq:LS} and \eqref{eq:Dxi} to
\begin{align}
D_\mu M & = \partial_\mu M - i {\mathcal L}_\mu M + i M {\mathcal R}_\mu\,,
\notag\\
D_\mu \xi_R & = \partial_\mu \xi_R - i V_\mu \xi_R + i \xi_R {\mathcal R}_\mu\,,
\notag\\
D_\mu \xi_L & = \partial_\mu \xi_L - i V_\mu \xi_L + i \xi_L {\mathcal L}_\mu\,,
\label{eq:DMext}
\end{align}
while the ordinary derivative in the nucleon kinetic term of Eq.~\eqref{eq:LN} becomes, for the mirror assignment of Eq.~\eqref{eq:mirror},
\begin{align}
D_\mu \psi_{1r} & = \left( \partial_\mu - i {\mathcal R}_\mu - i {\mathcal V}^{(0)}_\mu \right) \psi_{1r}\,,
\notag\\
D_\mu \psi_{1l} & = \left( \partial_\mu - i {\mathcal L}_\mu - i {\mathcal V}^{(0)}_\mu \right) \psi_{1l}\,,
\notag\\
D_\mu \psi_{2r} & = \left( \partial_\mu - i {\mathcal L}_\mu - i {\mathcal V}^{(0)}_\mu \right) \psi_{2r}\,,
\notag\\
D_\mu \psi_{2l} & = \left( \partial_\mu - i {\mathcal R}_\mu - i {\mathcal V}^{(0)}_\mu \right) \psi_{2l}\,,
\label{eq:Dpsi}
\end{align}
each chirality component carrying the field appropriate to its transformation law under Eq.~\eqref{eq:mirror}. The baryon and isospin chemical potentials enter as the time components
\begin{equation}
{\mathcal V}^{(0)}_\mu = \mu_B\, \delta_{\mu 0}\,, \qquad
{\mathcal L}_\mu = {\mathcal R}_\mu = \mu_I\, T_3\, \delta_{\mu 0}\,,
\label{eq:extmu}
\end{equation}
so that $\mu_p = \mu_B + \mu_I/2$ and $\mu_n = \mu_B - \mu_I/2$ for $T_3 = \tfrac{1}{2}{\rm diag}(1,-1)$. The thermodynamic potential is then
\begin{equation}
\Omega\left( \{\mu_i\}; \phi \right) = \sum_{i} \Omega_{\rm kin}\!\left(\mu_i^\ast, m_i^\ast\right) + \overline V\!\left( \phi \right),
\label{eq:Omega}
\end{equation}
with $\mu_i^\ast = \mu_i - \Sigma_{\tau_3(i)}(\phi)$, $\Omega_{\rm kin}$ the free-Fermi-gas contribution, and $\overline V$ of Eq.~\eqref{eq:Vbar}. The mean fields follow from $\partial\Omega/\partial\phi_a = 0$ at fixed $\{\mu_i\}$.

The two constructions coincide. Using $\partial\Omega_{\rm kin}/\partial\mu_i^\ast = -n_i$, $\partial\Omega_{\rm kin}/\partial m_i^\ast = n_{s,i}$ and $\partial\mu_i^\ast/\partial\phi_a = -\partial\Sigma_{\tau_3(i)}/\partial\phi_a$,
\begin{align}
\frac{\partial\Omega}{\partial\phi_a}\bigg|_{\{\mu_i\}}
& = \sum_i \left[ n_i\, \frac{\partial\Sigma_{\tau_3(i)}}{\partial\phi_a}
+ n_{s,i}\, \frac{\partial m_i^\ast}{\partial\phi_a} \right]
\notag\\
& \quad + \frac{\partial\overline V}{\partial\phi_a}
= \frac{\partial\varepsilon}{\partial\phi_a}\bigg|_{\{n_i\}}\,,
\label{eq:equiv}
\end{align}
where the last equality follows from Eq.~\eqref{eq:eps}. In the canonical formulation the vector source comes from the term $\sum_i \Sigma_{\tau_3(i)}n_i$ in $\varepsilon$; in the grand-canonical formulation it comes from the $\phi$ dependence of $\mu_i^\ast$. Equations~\eqref{eq:statchi}--\eqref{eq:statrho} are therefore the stationarity conditions of either functional, with the common scalar source terms $n_{s,i}\,\partial m_i^\ast/\partial\phi_a$.

The two potentials are Legendre transforms of one another, $\varepsilon = \Omega + \sum_i \mu_i n_i$, whence $\mu_i = \partial\varepsilon/\partial n_i = \mu_i^\ast + \Sigma_{\tau_3(i)}$ and
\begin{equation}
P = -\Omega = \sum_i \mu_i n_i - \varepsilon\,,
\label{eq:Peuler}
\end{equation}
the second form being the one evaluated in this work.

Two practical remarks should be made. First, beta equilibrium is imposed afterwards as $\mu_n = \mu_p + \mu_e$ (Sec.~\ref{sec:fit}). Second, the NJL table is given as a function of $\mu_B$, while our hadronic EoS is given as a function of the densities. The two are joined as functions of $n_B$ (Appendix~\ref{app:repro}), and Eq.~\eqref{eq:Peuler} relates the two descriptions.

\bibliography{sPDM-dilaton-2026Refs}

\end{document}